\documentclass[10pt,letterpaper]{article}
\usepackage[top=0.85in,left=2.75in,footskip=0.75in]{geometry}

\usepackage{graphicx}    
\usepackage{epsfig}
\usepackage{epstopdf}	 %needed for convertinf eps files to pdf
\usepackage{caption}
\usepackage{float}

\usepackage{amsmath,amssymb}

\usepackage{changepage}

\usepackage{textcomp,marvosym}

\usepackage{cite}

\usepackage[right]{lineno}

\usepackage{array}

\newcolumntype{+}{!{\vrule width 2pt}}

\newlength\savedwidth

\raggedright
\usepackage[aboveskip=1pt,labelfont=bf,labelsep=period,justification=raggedright,singlelinecheck=off]{caption}

\makeatletter
\renewcommand{\@biblabel}[1]{\quad#1.}
\makeatother

\usepackage{lastpage,fancyhdr,graphicx}
\usepackage{epstopdf}
\fancyheadoffset[L]{2.25in}
\fancyfootoffset[L]{2.25in}
\begin{document}
\vspace*{0.2in}

\begin{flushleft}
{\Large
\textbf\newline{Multicellular actomyosin cables in epithelia under external anisotropic stress} % Please use "sentence case" for title and headings (capitalize only the first word in a title (or heading), the first word in a subtitle (or subheading), and any proper nouns).
}
\newline
\\
Meryl A Spencer\textsuperscript{1},
Jesus Lopez-Gay \textsuperscript{2},
Hayden Nunley \textsuperscript{1},
Yohanns Bella\"{\i}che \textsuperscript{2},
David K Lubensky \textsuperscript{1}
\\
\bigskip
\textbf{1} Department of physics, University of Michigan Ann Arbor, Ann Arbor, MI, United States \\
\textbf{2} Institut Curie, 26 rue d'Ulm, 75248 Paris Cedex 05, France
\\
\bigskip

% Use the asterisk to denote corresponding authorship and provide email address in note below.
* dkluben@umich.edu

\end{flushleft}
\section*{Abstract}
The alignment of cell-cell junctions and associated cortical actomyosin across multiple cells to form supracellular cables in an epithelium is an example of the long range tissue organization that drives morphogenesis. Here we demonstrate that the ability of tissues to assemble these parallel cables depends on the initial packing topology of the cells in the epithelium. Using a computational vertex model we develop two methods of measuring a disordered tissue's favorability to forming cables under an external stress. These measures quantify the deformation of cells and the distribution of tension in the tissue under stress. Using these measures we show that passive stress-induced cell flow reduces a tissue’s ability to form cables, whereas oriented divisions create a packing which can sustain multiple parallel cables. These measures are applied to a region of the the {\it Drosophila} demonstrating a shift to a more cable-friendly packing after a wave of oriented divisions in the region. 

% Please keep the Author Summary between 150 and 200 words
% Use first person. PLOS ONE authors please skip this step. 
% Author Summary not valid for PLOS ONE submissions.   
\section*{Author summary}
Parallel actomyosin cables along cell-cell junctions are present in some epithelial tissues under anisotropic external mechanical stress. Here we demonstrate that the ability of tissues to form these parallel cables depends on the initial orientation of the cells in the epithelium. We devise two different measures of the ability of a tissue to form cables along a specified axis of high stress in disordered tissues. The first describes the percent of concave cells after simulated pulling on the tissue. The second describes the degree of force chain branching in the tissue after the same simulation. Using these measures we show that passive cell flow induced by anisotropic stress decreases the ability of the tissue to form cables, whereas oriented cell divisions along the axis of high stress move the cells into an orientation favorable for forming cables. We use our new measures to show that the {\it Drosophila} pupal epithelium changes topology to become more cable-friendly during development.

%\linenumbers

\section*{Introduction}
The long range organization and patterning of cells in an epithelium helps drive the morphogenesis of tissues \cite{FrancoisNat2007, JulicherCurrb2007, BellaicheScience2012, Osterfield}. One such pattern is the alignment of the junctions and associated cortex of multiple cells to form a continuous assembly of actomyosin, which we refer to as a cable (Fig~\ref{CFO}A). The appearance and role of isolated cables, for example, at compartment boundaries, and during wound healing and dorsal closure, has been well studied \cite{Osterfield,Roper2013, Michel2016, purse, West2017}. However, the role of multiple parallel cables within a single tissue in morphogenesis remains less well understood. A number of different tissues contain such parallel actomyosin cables, including the the {\it Drosophila} wing imaginal disk \cite{TaponEmbo2013, LecuitDev2013}, pupal wing blade \cite{IshiharaDev2013, EatonSemcdb2017}, and ventral epidermis \cite{DiNardoDev2010}, as well as the mouse heart \cite{KellyNatCom2017}. These cables are presumed to  have large contractile tensions, and thus to maintain anisotropic mechanical stresses. Although there is some knowledge of the molecular events driving the formation of these cables in some cases, little attention has been paid to the role of cell packing in aligning these junctions. Here we begin the work of understanding what role cell packing topology plays in cable formation.

\begin{figure}[t]
\begin{adjustwidth*}{-0in}{-2in}
{\includegraphics[width=1.3\textwidth]{Fig1-eps-converted-to.pdf}}
\caption{{\bf Cableness in ordered and disordered epithelia.}
Here and throughout we assume that a tensile stress is applied vertically to tissues. \\
{\bf A:} Cartoon of an epithelium with many cables highlighted in blue. We define a cable as a continuous, approximately straight line of myosin enriched junctions. \\
{\bf B:} Cartoon of tissues in the cable-forming orientation (CFO) and non-cable-forming orientation (NCFO). Edge color represents the tension on the edge; darker indicates higher tension. Cells in the CFO form brick shapes as they are stressed and reach a point at which their shape does not change further with increasing stress anisotropy. Cells in the NCFO collapse as they are stressed; the high tension edges are not neighbors, and do not form cables. \\
{\bf C:} Cartoon of epithelia with a range of cableness in the vertical direction. The packings with high cableness form many parallel cables when they are stressed, whereas the packings with low cableness form few to no cables. We define a {\it cably} packing to be one with high cableness.}
\label{CFO}
\end{adjustwidth*}
\end{figure}

An actomyosin cable forms when the junctional actomyosin across multiple cells aligns and assembles into a supra-cellular structure. It is currently unclear exactly how this continuity is achieved, although it is assumed that modified adherens junctions could link the ends of cables between adjacent cells, allowing forces to be transmitted along the cable \cite{Roper2015}. Actomyosin cables were first identified in wound healing where they were called `actin purse-strings' \cite{purse}. They have since been discovered in a number of different systems \cite{Roper2013}. There appear to be different pathways involved in cable formation that are system dependent. For example, the planar cell polarity proteins Frizzled and Flamingo are upstream activators of actin and myosin on the cable at the leading edge of dorsal closure in {\it Drosophila}, whereas Notch signaling is the upstream activator of cable formation at the dorsal-ventral compartment boundary  \cite{Chang2011, Kaltschmidt2002, Major2006}. Independently of the nature of the signaling pathway contributing to their formation, cables are defined by the up-regulation of myosin leading to increased tension along the affected junctions \cite{Roper2013,Landsberg2009,LecuitDev2013}. If the increase in tension is in {\it response} to an externally imposed mechanical stress, the cables are relatively stable, as in the pupal wing blade \cite{IshiharaDev2013}. If the stress anisotropy is instead internally generated, the cables are short lived, collapsing to form multicellular rosettes and driving tissue flow, as in convergent extension and neural tube closure \cite{ZallenJCB2014, Nishimura2012}. 

There are many interesting questions to ask about the relationship between  mechanical stress and tissue topology \cite{Vasquez2016, Wyatt2016, Noll2017, Bergmann2018}. When an external stress is applied to an epithelium it may react by selectively increasing its internal tension at the junctions and associated cortex, by up-regulating myosin along the axis of stress \cite{ZallenDevCell2009, SalbreuxTrends2012, LecuitTrends2012, MartinJOCB2014, Choi2016}. Here we show that this up-regulation will lead to the formation of parallel cables {\it only} when the initial cell arrangement is favorable. This is easy to see in the cartoons in Fig.~\ref{CFO} where we show two perfectly ordered tissues in the `cable forming orientation' (CFO) and the `non-cable forming orientation' (NCFO). These two orientations behave differently under stress. As the stress anisotropy increases, cells in the CFO become brick shaped and cables from along the columns of cells. Because the brick-like packing contains lines of perfectly vertical junctions, cells in the CFO can in principal support any vertical stress without collapsing (provided, of course, that they can upregulate the tension on the cables sufficiently). Cells in the NCFO, in contrast, become highly elongated under an imposed stress anisotropy; they eventually collapse down to zero area as the applied stress increases. These cells are unable to form cables, because their high tension vertical edges cannot align unless the cells shrink to nothing. Previous studies have used the number of neighboring high tension vertical edges to identify cables \cite{ZallenDevCell2009}.  

Importantly, although the CFO and NCFO are related by an overall $30^o$ rotation, most epithelia are constrained at their boundaries such that they cannot rigidly rotate. Thus, shifting between the CFO and NCFO requires changing the cellular packing topology. In this paper we focus on the formation of multiple parallel cables, as opposed to single isolated cables, such as those that form at compartment boundaries \cite{Roper2013, Sussman2018}. For the duration of the paper we always define the vertical axis to have higher stress, such that cables only form parallel to the $y$-axis. 

Although the difference between the CFO and the NCFO can easily be understood from a few pictures, it is far less obvious how this distinction translates to a more realistic, disordered cell packing that is intermediate between the two limiting cases.  Our ultimate aim is to understand how cell packing topology affects cable formation in just such disordered tissue. As a first step towards this larger goal, here we devise ways of quantifying how favorable a tissue topology is to forming cables in a given direction, a quality which we will refer to as {\it cableness}. Fig.~\ref{CFO}C shows cell packings we would intuitively like to define as having different cableness in both their stress free and anisotropically stressed states.  Those with high cableness form many parallel cables under anisotropic stress, whereas those with low cableness form few to no cables. Additionally, the cells in tissues with low cableness are more likely to become highly deformed under stress than cells in tissues with high cableness.  In tissues with high cableness, the high tension edges align to form continuous cables, so that the force is evenly distributed through the tissue. In tissues with low cableness, most high tension edges are not connected to one another. Few cables form and lines of force frequently branch. Below we translate these qualitative ideas into quantitative measures of cableness. We define cableness to be a measure of a cell packing topology's {\it potential} to form cables under applied stress, so that a tissue has the same cableness before and after an applied stress (assuming it does not change topology). Therefore, the cableness of a tissue is a function solely of its topology and not of the presence or absence of cables at any given time (Fig.~\ref{CFO}C).

Throughout this paper we will base our understanding of mechanical forces in an epithelium on the vertex model framework \cite{SpencerEpje2017, GavaghanProg2013, ShvartsmanBioj2014}. We will begin with a brief description of our computational implementation of the vertex model, including some relevant structural choices we made. We will then define a {\it stretching procedure} to computationally add anisotropic stress to tissues, which we will use to determine their cableness. Next we define two different cableness measures based on the geometry and distribution of tension in stretched tissues. Using these two measures we will investigate how the cableness of a tissue changes as cells undergo both oriented T1 topological changes \cite{SpencerEpje2017} and oriented divisions. Finally, we will apply our measures to data taken from the {\it Drosophila} pupa notum, where we will find that the cell arrangement changes over time to increase the tissue's cableness in response to an applied stress.

%%%%%%%%%%%%%%%%%%%%%%%%%%%%%%%%%%%%%%%%%%%%%%%%%%%%%%%%%%%%%%%%%%%%%%%%%%%%%%%%%%%%%%%%%%%%%%%%%%%%%%%%%%%%%%%%%%%%%%%%%%%%%%%%%%%%%%%%%%%%%%%%%%%%%%%%%%%%%%%%%%%%%%%%%%%%%%%%%%%%%%%%%%%%%%%%%%%%%%%%%%%%%%%%%%%%%%%%%%%%%%%%%%%%%%%%%%%%%%%%%%%%%%%%%%%%%%%%%%%%%%%%%%%%%%%%%%%%%%%%%%%%%%%%%%%%%%%%%%%%%%%%%%%%%%%%%%%%%%%%%%%%%%%%%%%%%%%%%%%%%%%%%%%%%%%%%%%%%%%%%%%%%%%%%%%%%%%%%%%%%%%%%%%%%%%%%%%%%%%%%%%%%%%%%%%%%%%%%%%%%%%%%%%%%%%%%%
\section*{Materials and methods}
In this section we will  give a description of the vertex model, which is the theoretical and computational model we use to understand the physics of an epithelium. There are many different variations of the vertex model, each of which has different properties and is best used in different situations. In this section we define the variants of the vertex model used in the paper; subsequent sections explain why we've chosen one or the other. We also define the stretching procedure which we use when calculating a tissue's cableness. 

%%%%%%%%%%%%%%%%%%%%%%%%%%%%%%%%%%%%%%%%%%%%%%%%%%%%%%%%%%%%%%%%%%%%%%%%%%%%%%%%%%%%%%%%%%%%%%%%%%%%%%%%%%%%%%%%%%%%%%%%%%%%%%%%%%%%%%%%%%%%%%%%%%%%%%%%%%%%%%%%%%%%%%%%%%%%%%%%%%%%%%%%%%%%%%%%%%%%%%%%%%%%%%%%%%%%%%%%%%%%%%%%%%%%%%%%%%%%%%%%%%%%%%%%%%%%%%%%%%%%%%%%%%%%%%%%%%%%%%%%%%%%%%%%%%%%%%%%%%%%%%%%%%%%%%%%%%%%%%%%%%%%%%%%%%%%%%%%%%%%%%%%%%%%%%%%%%%%%%%%%%%%%%%%%%%%%%%%%%%%%%%%%%%%%%%%%%%%%%%%%%%%%%%%%%%%%%%%%%%%%%%%%%%%%%%%%%
\subsection*{Theoretical framework: vertex model}
Vertex models are a common way of understanding the physics of simple epithelia at the level of cellular scale mechanical forces \cite{JulicherCurrb2007, SpencerEpje2017, VonMeringPloscb2011, GavaghanProg2013, ShvartsmanBioj2014}. They describe an epithelium as a quasi-two-dimentional sheet composed of cells, edges, and vertices. Edges describe the cortex at the level of the adherens junctions. A vertex is defined as any place three or more edges meet. Both cells and edges push and pull on vertices through mechanical forces. The mechanical force on a vertex is given by
\begin{equation}
\vec{F}=\sum_{ \text{edges } i} {\gamma_i\hat{l}_i}+ \sum_{\text{cells } j}{\frac{P_j}{2} \hat{z}\times\big(\vec{l}_{j1}-\vec{l}_{j2}\big)},
\label{eq:1}
\end{equation}
where $\vec{l}$ represents the length and orientation of an edge pointing out from the vertex, and $\hat{z}$ is perpendicular to the plane of the epithelium \cite{SpencerEpje2017}. The first term describes the force from the edges on the vertex. The sum runs over all edges connected to the vertex. The strength of the force on each edge is given by its tension $\gamma_i$. The second term describes the force from the vertex's neighboring cells. Each cell has a pressure $P_j$ determined by its deformation from its preferred area and acting in the direction of $\hat{z}\times\big(\vec{l}_{j1}-\vec{l}_{j2}\big)$ where $\vec{l}_{j1}$ and $\vec{l}_{j2}$ are the edges of cell $j$ adjacent to the vertex taken clockwise. Vertex motion is assumed to be over-damped so that the velocity $\vec{v}_k$ of a vertex $k$ is proportional to the force acting on it. 
\begin{equation}
\alpha \vec{v_k}=\vec{F}_{k},
\label{eq:mechmotion}
\end{equation}
where $F_{k}$ is the force on vertex $k$ given by Eq~\ref{eq:1}. From some initial placement of cells, edges, and vertices the equation of motion for every vertex is integrated forward in time to model the dynamics of the tissue. 

%%%%%%%%%%%%%%%%%%%%%%%%%%%%%%%%%%%%%%%%%%%%%%%%%%%%%%%%%%%%%%%%%%%%%%%%%%%%%%%%%%%%%%%%%%%%%%%%%%%%%%%%%%%%%%%%%%%%%%%%%%%%%%%%%%%%%%%%%%%%%%%%%%%%%%%%%%%%%%%%%%%%%%%%%%%%%%%%%%%%%%%%%%%%%%%%%%%%%%%%%%%%%%%%%%%%%%%%%%%%%%%%%%%%%%%%%%%%%%%%%%%%%%%%%%%%%%%%%%%%%%%%%%%%%%%%%%%%%%%%%%%%%%%%%%%%%%%%%%%%%%%%%%%%%%%%%%%%%%%%%%%%%%%%%%%%%%%%%%%%%%%%%%%%%%%%%%%%%%%%%%%%%%%%%%%%%%%%%%%%%%%%%%%%%%%%%%%%%%%%%%%%%%%%%%%%%%%%%%%%%%%%%%%%%%%%%%
\subsubsection*{Choices in the vertex model}
\begin{figure}[t]
\begin{adjustwidth*}{-0in}{-2in}
{\includegraphics[width=1.3\textwidth]{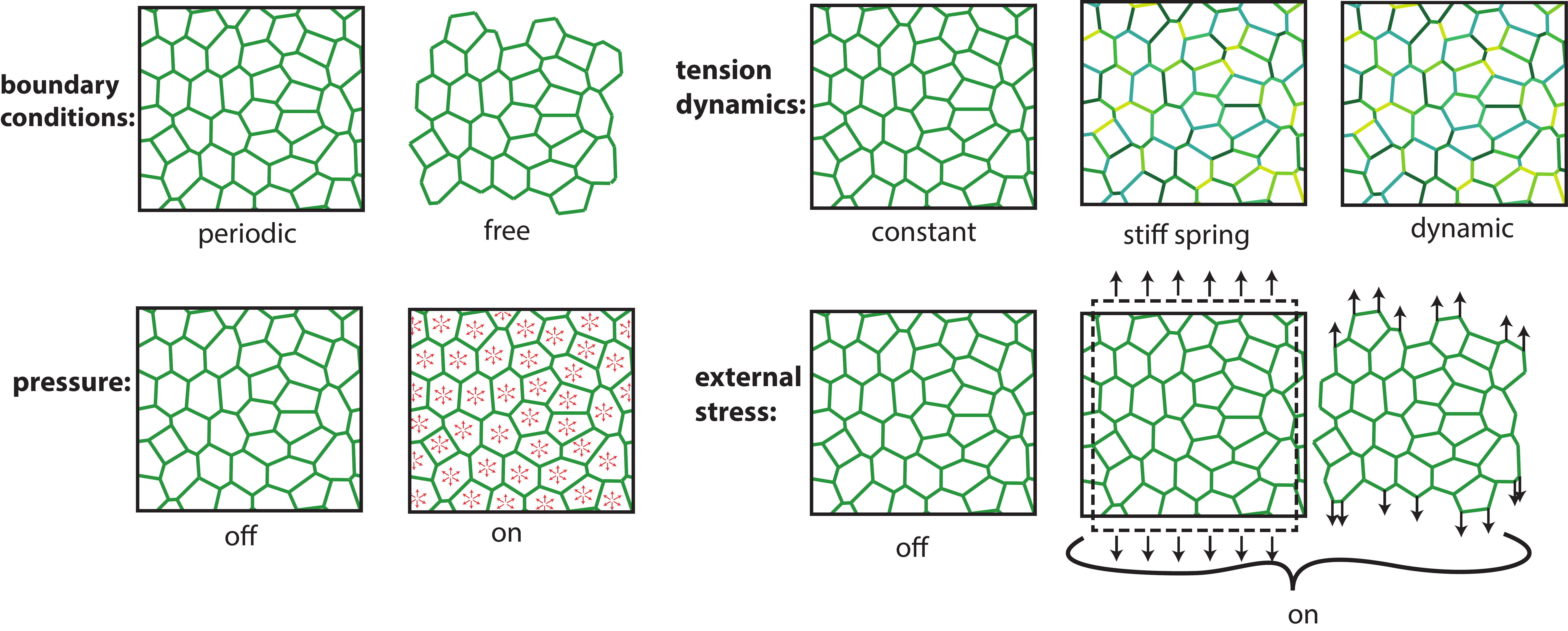}}
\caption{{\bf Vertex model options.}
Cartoon of some of the options available to customize a vertex model. The tissue can have free or periodic boundary conditions. Cells may or may not have pressure, and edge tensions may be constant and identical for all edges or evolve according to different rules. Different colored edges in the cartoon represent edges with different tensions. The tissue can experience anisotropic stresses or not. As discussed in the text, anisotropic stress can be implemented in different ways. Each of these choices can be made independently of the others.}
\label{VMChoices}
\end{adjustwidth*}
\end{figure}

There are a number of choices to be made when implementing a vertex model. Because we use many different forms of the vertex model in this paper, we will briefly cover the choices that need to be made and their implications. These choices are summarized in Fig.~\ref{VMChoices}. 

{\bf Choice 1: boundary conditions} We use two different boundary conditions, free and periodic. When using free boundary conditions the equations of motion of the vertices are given by Eq \ref{eq:mechmotion}. The major upside to using free boundary conditions is that the vertex model can be seeded using skeletonizations of experimental images. The downside is that one must specify the behavior of boundary cells. When using periodic boundary conditions the cells live in a rectangular box of size $L_x$ by $L_y$. One could hold $L_x$ and $L_y$ fixed, but we generally want to be in a constant tension ensemble so we let the box size evolve in time according to the algorithms in  S1 Appendix. In the limit of a large tissue we expect the boundary conditions to be irrelevant.

{\bf Choice 2: cell pressure} Vertex models may or may not include pressure forces. Including cell pressures gives a more realistic simulation, and pressures are essential when edge tensions are contractile. However, pressures can be dispensed with in other contexts, which generally decreases simulation runtime. 

{\bf Choice 3: form of edge tension} Every edge has some tension $\gamma_i$. We often assume that every edge has the same tension $\gamma$, which corresponds to the case in which edge tension is slow to respond to imposed stress. We will also use length dependent edge tensions so that edges resist compression and expansion while rotating freely. When edges respond quickly to stress we choose a simple form of 
\begin{equation}
\beta_\gamma \dot{\gamma_i}= \big(\vec{F}_{a} - \vec{F}_b\big) \cdot{\hat{l}_i}, 
\end{equation} 
where $\vec{F}_a$ and $\vec{F}_b$ are the forces on the vertices of edge $i$, as in \cite{Shraiman17}. We also sometimes allow the tension on every edge to act like a stiff spring, so that
\begin{equation}
\gamma=\kappa(l-l_0).
\end{equation} 

{\bf Choice 4: external stresses.} We generally want to set stresses in the simulations and let the tissue size vary based on this applied stress. When including external stress on a tissue with periodic boundary conditions the box size changes according to the difference between internal and external stress, as described in S1 Appendix. When including external stresses in tissues with free boundary conditions a constant force is applied to every boundary vertex, which we take to be any vertex with only two edges. The same total force is applied to the top and bottom row of vertices. We distribute this force evenly amongst all the vertices on one boundary to prevent large torques. We choose this procedure, rather than introducing a rigid boundary to which we apply a force, because we want to ensure that stress is reasonably evenly distributed across the tissue. 

%%%%%%%%%%%%%%%%%%%%%%%%%%%%%%%%%%%%%%%%%%%%%%%%%%%%%%%%%%%%%%%%%%%%%%%%%%%%%%%%%%%%%%%%%%%%%%%%%%%%%%%%%%%%%%%%%%%%%%%%%%%%%%%%%%%%%%%%%%%%%%%%%%%%%%%%%%%%%%%%%%%%%%%%%%%%%%%%%%%%%%%%%%%%%%%%%%%%%%%%%%%%%%%%%%%%%%%%%%%%%%%%%%%%%%%%%%%%%%%%%%%%%%%%%%%%%%%%%%%%%%%%%%%%%%%%%%%%%%%%%%%%%%%%%%%%%%%%%%%%%%%%%%%%%%%%%%%%%%%%%%%%%%%%%%%%%%%%%%%%%%%%%%%%%%%%%%%%%%%%%%%%%%%%%%%%%%%%%%%%%%%%%%%%%%%%%%%%%%%%%%%%%%%%%%%%%%%%%%%%%%%%%%%%%%%%%%
\subsection*{Stretching procedure}
\begin{figure}[h]
\begin{adjustwidth*}{-0in}{-2in}
{\includegraphics[width=1.3\textwidth]{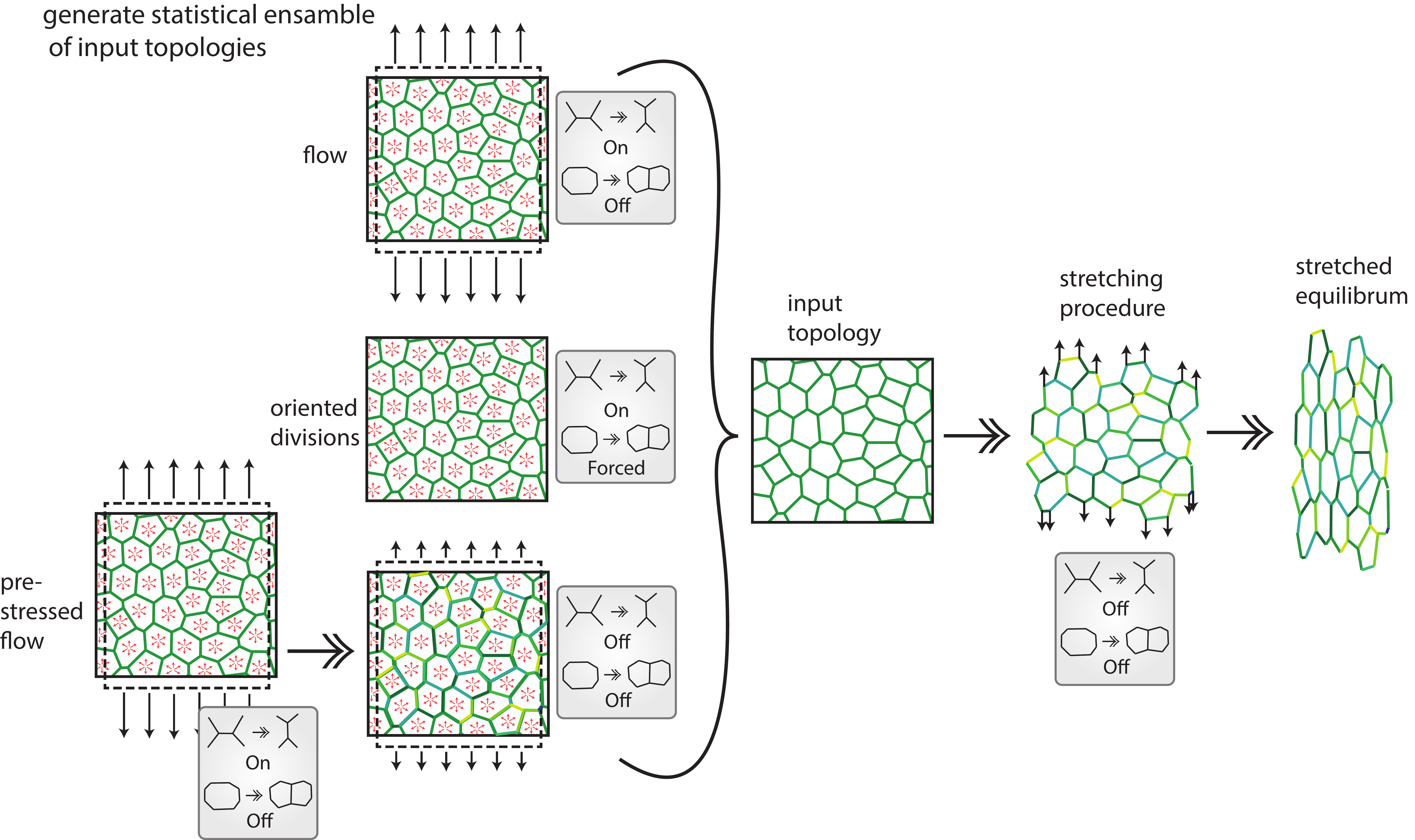}}
\caption{{\bf Stretching procedure.}
Cartoon of the major vertex model options used to investigate cableness. Statistical ensembles of input topologies are created in three different ways. Flow topologies are created by exerting a large anisotropic stress on packings with pressure effects and constant edge tensions. No cell divisions are allowed and T1s happen when energetically favorable. Oriented division topologies are created by dividing each cell exactly once in a tissue with pressure effects and constant edge tensions. T1s happen when energetically favorable. Pre-stressed flow topologies are seeded with the topology created by the flow procedure. A small anisotropic stress is applied to the tissue, which has pressure effects and dynamically changing edge tensions. Neither divisions nor T1s are allowed. The {\it stretching procedure} is a vertex model where anisotropic stress is applied to a tissue with free boundary conditions, no pressure effects, and stiff-spring edge tensions. We refer to this final state of mechanical equilibrium after stretching as the {\it stretched equilibrium}. All topological changes are suppressed. The resulting stretched equilibrium packing is used to measure cableness. 
}
\label{VMUsed}
\end{adjustwidth*}
\end{figure}

In order to develop measures of cableness we will look at the behavior of disordered cell packings under highly anisotropic stress.  We thus want a standardized way to impose anisotropic stress that captures whether a given packing can align its edges into cables.  To this end, we introduce a \textit{stretching procedure} that allows the edges in any cell packing to rotate and increase their tension in a simple way in response to a strong, uniaxial stress while maintaining a constant topology.  We expect the behavior of cells in this model to qualitatively but not quantitatively reflect behavior under more involved rules more directly inspired by specific biological systems.  Fig.~\ref{VMUsed} shows the variants of the vertex model we used. First an input topology is created through a variety of methods which we will discuss later. Then the stretching procedure is applied to the input topology. Free boundary conditions were chosen so that the method can be easily applied to skeletonizations of experimental images of epithelia. Cells do not exert any pressure on their surrounding vertices; this cuts down on the number of free parameters in the model and helps to exaggerate cell deformations. We expect that the pressure does not play a large role in the real behavior of cells highly anisotropic under stress, as isotropic cell pressures can never counteract anisotropic stresses. The tension on every edge acts like a stiff spring:
\begin{equation}
\gamma=\kappa(l-l_0),
\end{equation} 
where $\kappa$ represents the spring constant and $l_0$ is the initial length of the edge. We make this choice because it is the simplest version of feedback on the edge tension in response to mechanical stretching. A fixed, external stress is applied and the tissue relaxes to its equilibrium state which we refer to as the {\it stretched} state. We do not allow any topological changes since we are looking to measure the cableness of the initial topology. When looking at properties of the stretched state we always restrict measures to the middle 50\% of cells to avoid any boundary effects.  A network of threefold-coordinated vertices without any cell pressures is generically under-constrained, and we expect it to be able to undergo a finite deformation without generating any internal stress by rotating its edges at constant length.  At some point as it is stretched, it will undergo a transition from this floppy state to a stiff state where the vertices are aligned so as to support a nonzero tension in the edges \cite{Feng2016,Sharma2016,Vermeulen2017}.  The final state of mechanical equilibrium in our stretching procedure is above this stiffening transition.  In general, this state could depend on the magnitude of the stress we impose in a way that reflects the nonlinear network elasticity beyond the transition \cite{Storm2005,Ronceray2016}.  In practice, however, we use a large value of $\kappa$ (S3 Table) so that edge length depends very weakly on edge tension.  In this limit, the spatial arrangement of vertices in the final state of self stress is essentially independent of the imposed stress, which serves only to set an overall scale for the edge tensions.

%%%%%%%%%%%%%%%%%%%%%%%%%%%%%%%%%%%%%%%%%%%%%%%%%%%%%%%%%%%%%%%%%%%%%%%%%%%%%%%%%%%%%%%%%%%%%%%%%%%%%%%%%%%%%%%%%%%%%%%%%%%%%%%%%%%%%%%%%%%%%%%%%%%%%%%%%%%%%%%%%%%%%%%%%%%%%%%%%%%%%%%%%%%%%%%%%%%%%%%%%%%%%%%%%%%%%%%%%%%%%%%%%%%%%%%%%%%%%%%%%%%%%%%%%%%%%%%%%%%%%%%%%%%%%%%%%%%%%%%%%%%%%%%%%%%%%%%%%%%%%%%%%%%%%%%%%%%%%%%%%%%%%%%%%%%%%%%%%%%%%%%%%%%%%%%%%%%%%%%%%%%%%%%%%%%%%%%%%%%%%%%%%%%%%%%%%%%%%%%%%%%%%%%%%%%%%%%%%%%%%%%%%%%%%%%%%%
\section*{Results}
We have split our results into three sections. In the first section we will describe how we measured cableness in disordered tissues. In the second section we will use our new cableness measures to show that oriented cell divisions promote cableness. In the third section we will discuss data from the {\it Drosophila} pupa notum which suggests that the notum becomes more cably after a round of oriented cell divisions. 

\subsection*{Defining measures of cableness}
We would like to produce a measure of cableness that corresponds to our intuitive notion of cableness (e.g., Fig. \ref{CFO}C) and that can be applied not only to simulation results but also directly to skeletonized images of real cell packings in order to determine the cableness of real tissues. This means the measure should depend solely on edge orientation and length which can be determined from images. Additionally the measure should work on pre-stressed as well as unstressed cell packings, as tissues observed in experiments will not typically be stress-free. Recall that our definition of cableness depends on a tissue's potential to form cables and is therefore a function solely of its topology, not of the presence or absence of cables at any given time.

Before we look for a measure that will work on disordered tissues we will look for insights from our toy model tissues the CFO and NCFO (Fig.~\ref{CFO}B). An obvious difference between the CFO and NCFO is the edge orientation as measured by the average of $\cos(6\theta)$, where $\theta$ is the angle of the edge to the horizontal axis, which is 1 for the CFO and -1 for the NCFO. In fact this and closely related measures have been used before to quantify orientational order\cite{IshiharaDev2013, RN4, Chen2016}. However, this measure only works in stress free tissues, and does not correctly distinguish cableness in geometries under stress. For example $\langle\cos(6\theta)\rangle$ for the CFO under high stress anisotropy is $-1/3$. Moreover, it's not clear that $\langle\cos(6\theta)\rangle$ necessarily predicts with cableness in disordered packings. Below we define alternate metrics that correlate well with $\langle\cos(6\theta)\rangle$ in unstressed tissues but that behave better under stress and correspond more directly to intuitive notions of cableness.

\subsubsection*{Creating control packings through cell flow}
\begin{figure}[h]
\begin{adjustwidth*}{-0in}{-2in}
{\includegraphics[width=1.3\textwidth]{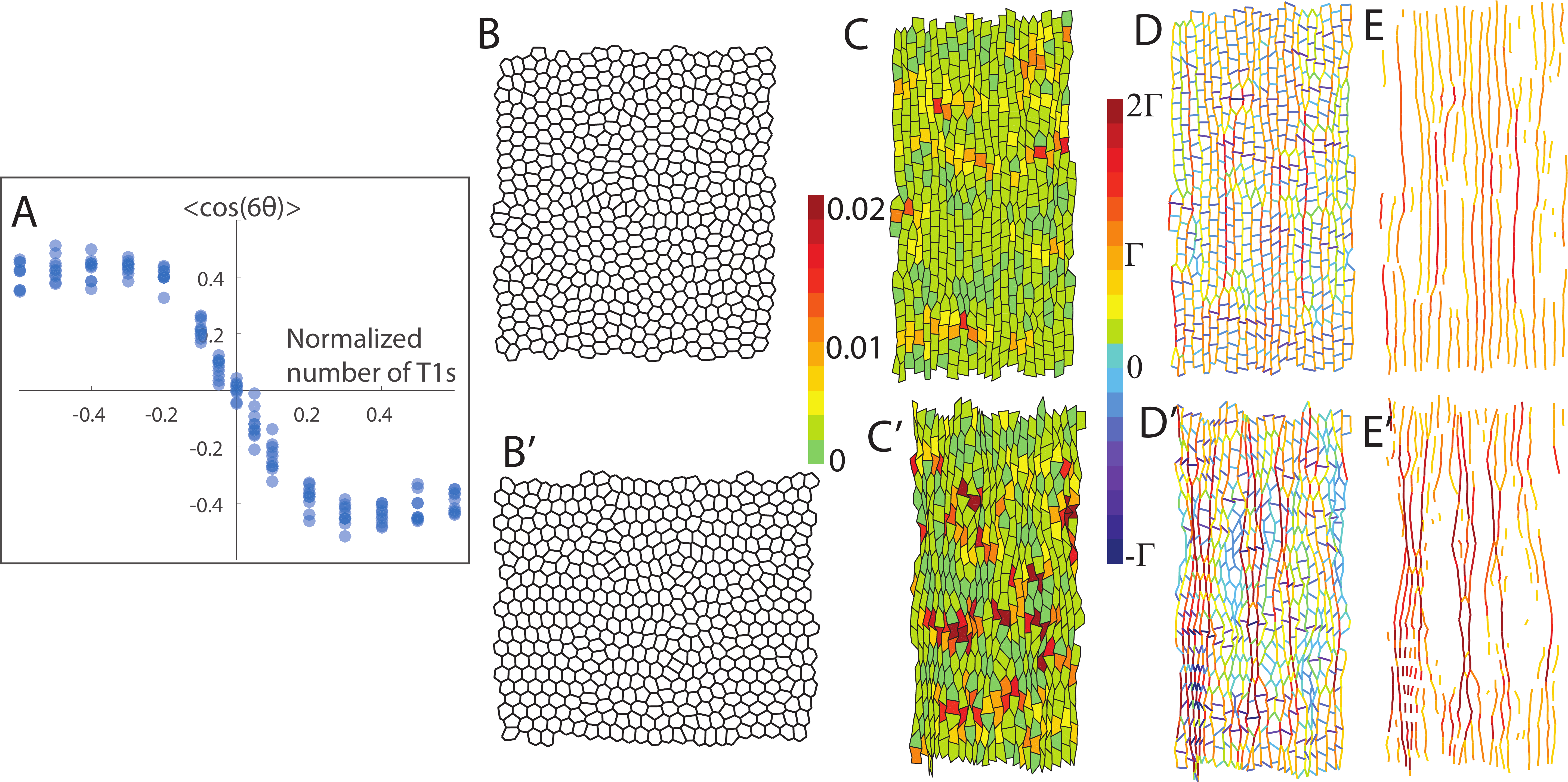}}
\caption{{\bf Cableness of flowing tissues.}\\
{\bf A} The average edge orientation of tissues as measured by the average of $\cos6\theta$ (where $\theta$ is the angle of the edge from the horizontal) is shown for our statistical ensemble of flow based packings. Each point represents data from the central region of one packing of 1000 cells (approximately 500 cells in the central region). Negative values on the $x$-axis indicate tissues that flowed perpendicular to the axis of cableness. \\
{\bf B (B')} Resulting packing from flowing cells perpendicular (parallel)  to the cableness axis with 60\% of edges undergoing a T1 transition. \\
{\bf C (C')} Stretched equilibrium packing, color indicates cell convexity as given by $(P-H)/H$, where $P$ is the perimeter and $H$ is the convex hull. \\
{\bf D (D')} Stretched equilibrium packing, color indicates edge tension, where $\Gamma$ is the average force applied to boundary vertices.  \\
{\bf E (E')} Same as D (D') with only high tension edges shown. 
}
\label{flow}
\end{adjustwidth*}
\end{figure}

As we move from our toy models to disordered tissues we need to generate a statistical ensemble of disordered packings that we have independent reason to believe are more or less cably. It is well established that flowing tissues experience oriented T1 topological transitions, in which an edge shrinks to a fourfold vertex and then grows in the perpendicular direction causing cells to exchange neighbors \cite{SpencerEpje2017} (S1 Figure). When edge tensions are constant tissues under large external stress anisotropy will flow \cite{LarsonChemPhy1995, Schmidt1995, Hess1999}. These oriented T1 transitions change the value of $\langle \cos(6\theta)\rangle$ in the tissue, and thus we expect the cableness to change as well (Fig.~\ref{flow}A). It is notable that vertical flow induces oriented T1 transitions from horizontal to vertical edges which pushes the tissue into a less cably direction. Therefore a tissue's passive response to stress inhibits the tissue's ability to form cables.  

To generate packings with varying cableness we thus induced flow in a vertex model. We measured the extent of the flow, and thus the expected cableness, by counting the number of T1 transitions. A cartoon of the vertex model we used to generate these packings is given in Fig.~\ref{VMUsed} in the panel labeled `flow'. The simulation is seeded with a Voronoi tessellation of randomly placed points in the plane and  relaxes from this initial condition under isotropic stress. Once it has relaxed an external anisotropic stress of $\sigma_{yy}=2\gamma\sqrt{N_c}/A$, $\sigma_{xx}=\gamma\sqrt{N_c}/(2A)$ (where $N_c$ is the number of cells and $A$ is the total area of the tissue) is applied and the tissue is allowed to flow until a specified number of T1 transitions is reached. Once the desired number of T1s has been reached the tissue again relaxes under isotropic stress. Periodic boundary conditions are used for convenience. Pressure effects from cells are turned on to give a more realistic simulation. Edge tensions are constant to allow for cells to flow.   We created tissues of 1000 cells each by inducing various levels of flow in both the vertical and horizontal directions. We then applied the stretching procedure.

\subsubsection*{Initial cableness measures}
In order to define measures of cableness that correspond to physical properties of tissues under stress, we began by looking at the qualitative behavior of our statistical ensemble of flowing tissues in their stretched state (Fig.~\ref{flow}B-E). The primed panels represent a tissue which flowed parallel to the axis of cableness, expected to be less cably, whereas unprimed frames correspond to a tissue that flowed perpendicular to the axis of cableness, expected to be more cably. 

Fig.~\ref{flow}C shows the convexity of cells in the stretched equilibrium as measured by the normalized difference between the perimeter and convex hull. There is a clear difference in the number of highly concave cells in the two tissues. Our first measure of cableness $\mathcal{C}$ is defined as 
\begin{equation}
\mathcal{C}= \frac{1}{N_{cells}}\sum_{cells} {\Theta \bigg(\frac{P-H}{H}-\epsilon \bigg)}, 
\end{equation}
where $\Theta$ is the Heaviside step function, $P$ is the perimeter, $H$ is the convex hull and $\epsilon$ is a small, positive cutoff. This essentially measures the fraction of concave cells in the tissue, disregarding cells that are only barely concave.  Thus, small values of $\mathcal{C}$ correspond to high cableness. We require the cutoff $\epsilon$ because the brick shaped cells in the highly cably packings are often slightly very slightly concave, to an extent that is neither apparent to the naked eye nor likely to be biologically meaningful. We use an $\epsilon$ of $0.01$ because it is roughly an order of magnitude larger than the numerical error of our simulation. 

A second measure of cableness comes from the distribution of tensions in the equilibrium state after pulling. Fig.~\ref{flow}E shows the difference in behavior of the load bearing edges between a cably and non-cably tissue. The tissue with higher cableness has more evenly spaced cables with moderate tensions, whereas the tissue with low cableness has few unevenly spaced cables with high tensions. We create a quantitative scalar measure $\mathcal{T}$ that describes the difference in the way the tension is distributed between cably and non-cably packings. 

\begin{figure}[t]
\begin{adjustwidth*}{-0in}{-2in}
{\includegraphics[width=1.3\textwidth]{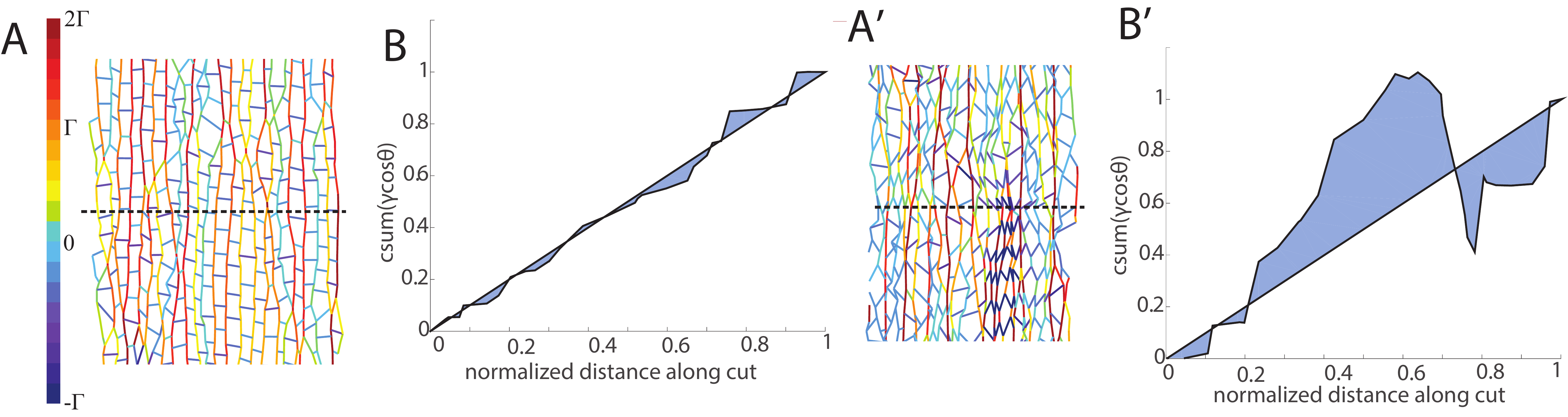}}
\caption{ {\bf Defining cableness measure $\mathcal{T}$.} \\
{\bf A, A'} A more and less cably tissue after applying the stretching procedure. Color indicates edge tension, where $\Gamma$ is the average force applied to boundary vertices.   \\
{\bf B, B'} Plot of the difference between the unit line and the normalized cumulative sum $h(x)$ as a function of normalized distance $x$ across the horizontal cut shown in A, A' (dashed line). The value of $\mathcal{T}$ is given by the area of the shaded region divided by the number of edges cut. \\
}
\label{defT}
\end{adjustwidth*}
\end{figure}

\begin{figure}[t]
\begin{adjustwidth*}{-0in}{-2in}
{\includegraphics[width=1.3\textwidth]{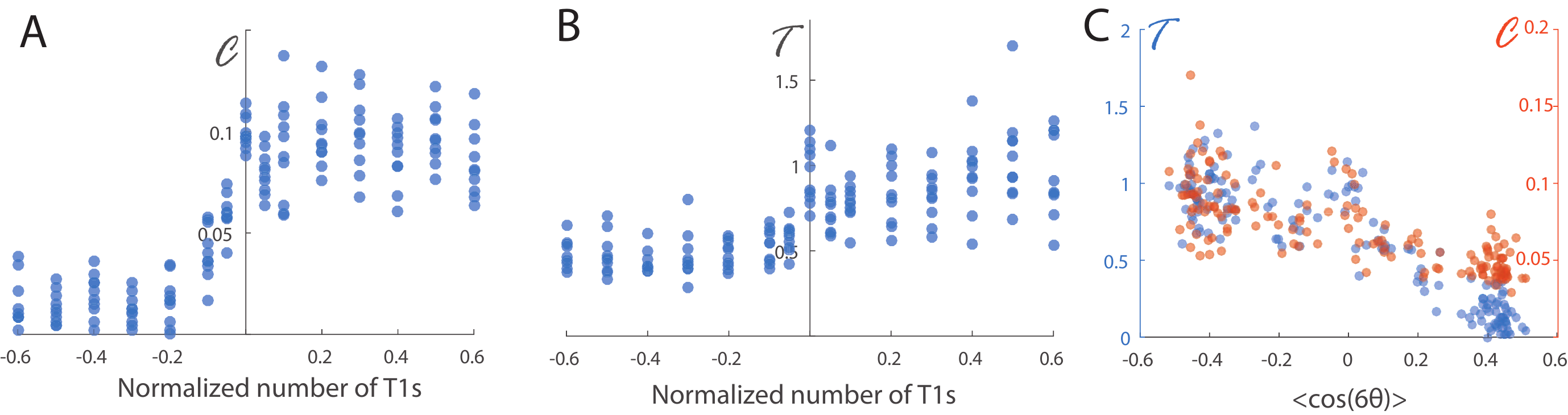}}
\caption{{\bf Cableness of flowing tissues.}\\
{\bf A} Result of applying the cell convexity based cableness measure $\mathcal{C}$ to the packings generated by flow. Each point represents data from the central region of one packing of 1000 cells (approximately 500 cells in the central region). Negative values on the $x$-axis indicate flow parallel to the cableness axis. \\
{\bf B} Result of applying the tension based cableness measure $\mathcal{T}$ to the packings generated by flow. Each point represents data from the central region of one packing of 1000 cells (approximately 500 cells in the central region). Negative values on the $x$-axis indicate flow parallel to the cableness axis. \\
{\bf C} Cableness measures $\mathcal{C}$ (red) and $\mathcal{T}$ (blue) as a function of the edge orientation. Smaller values of $\mathcal{C}$ and $\mathcal{T}$ correspond to higher cableness. 
}
\label{flowCT}
\end{adjustwidth*}
\end{figure}

To get $\mathcal{T}$ we look at a horizontal cut through the tissue as shown in Fig. \ref{defT}A(A$^\prime$). Let
\begin{equation}
h(x)= \frac{ \sum_0^x \gamma_i\cos\theta_i}{\sum_0^1 \gamma_i\cos\theta_i}
\end{equation}
be the normalized cumulative sum of the vertical component of the edge tension of every edge which intersects the cut up to normalized distance $x$ along the cut.  In a cably tissue, where there are many lines of high tension roughly evenly spaced, $h(x)$ will be approximately linear. In the non-cably tissue, where there are only a few lines of high tension in clusters throughout the tissue, $h(x)$ will be step-like. The integral of the difference between the unit linear function and $h(x)$ (blue shaded reigon in \ref{defT}B(B$^\prime$) ) should be low for cably packings and high for non-cably packings. We define $\mathcal{T}_h$ as
\begin{equation}
\mathcal{T}_h=\frac{1}{N} \int_0^1dx \sqrt{\big[x-h(x)\big]^2 }, 
\end{equation}
where $N$ is the number of edges intersecting the cut. In order to remove boundary effects from starting the cumulative sum $h(x)$ at the left edge of the tissue, we calculate $\mathcal{T}_h$ starting at every edge and wrapping around the tissue and always take the minimum value $\mathcal{T}_{h,\text{min}}$.  The cableness measure $\mathcal{T}$ is the median of $\mathcal{T}_{h,\text{min}}$ over 50 evenly spaced horizontal cuts through the tissue. 
Fig.~\ref{flowCT} shows the result of applying our two cableness measures to the topologies generated by flow. Both measures decrease with increasing $\langle \cos(6\theta)\rangle$, and are in good agreement with one another. However, the values of cableness at no flow are higher than we would expect from the overall trend. We will investigate the reason for this bump in the next section. 

\subsubsection*{Cableness along a single axis depends on the level of tissue disorder }

We want to understand why there is a bump in both of our cableness measures at no flow. One difference between the tissues which have minimal flow and tissues which do not flow is the level of disorder as measured by the standard deviation in the number of edges per cell, which has been used as a measure of topological order in tissues \cite{BaumDev2017}.  We created three ensembles of packings with$\langle \cos(6\theta)\rangle \simeq 0$, which we might naively expect to have the same cableness. The first was identical to the ensemble used as the initial conditions for the flow simulations (see previous section). The second ensemble was created by inducing random cell divisions in an initially isotropic packing. From an initially isotropic packing cells were chosen at random to divide along a random axis until every cell divided exactly once. To create the third ensemble we began with a hexagonal packing and randomly selected half of the vertical edges to undergo T1 transitions.

\begin{figure}[h]
\begin{adjustwidth*}{-2in}{-0in}
{\includegraphics[width=1.3\textwidth]{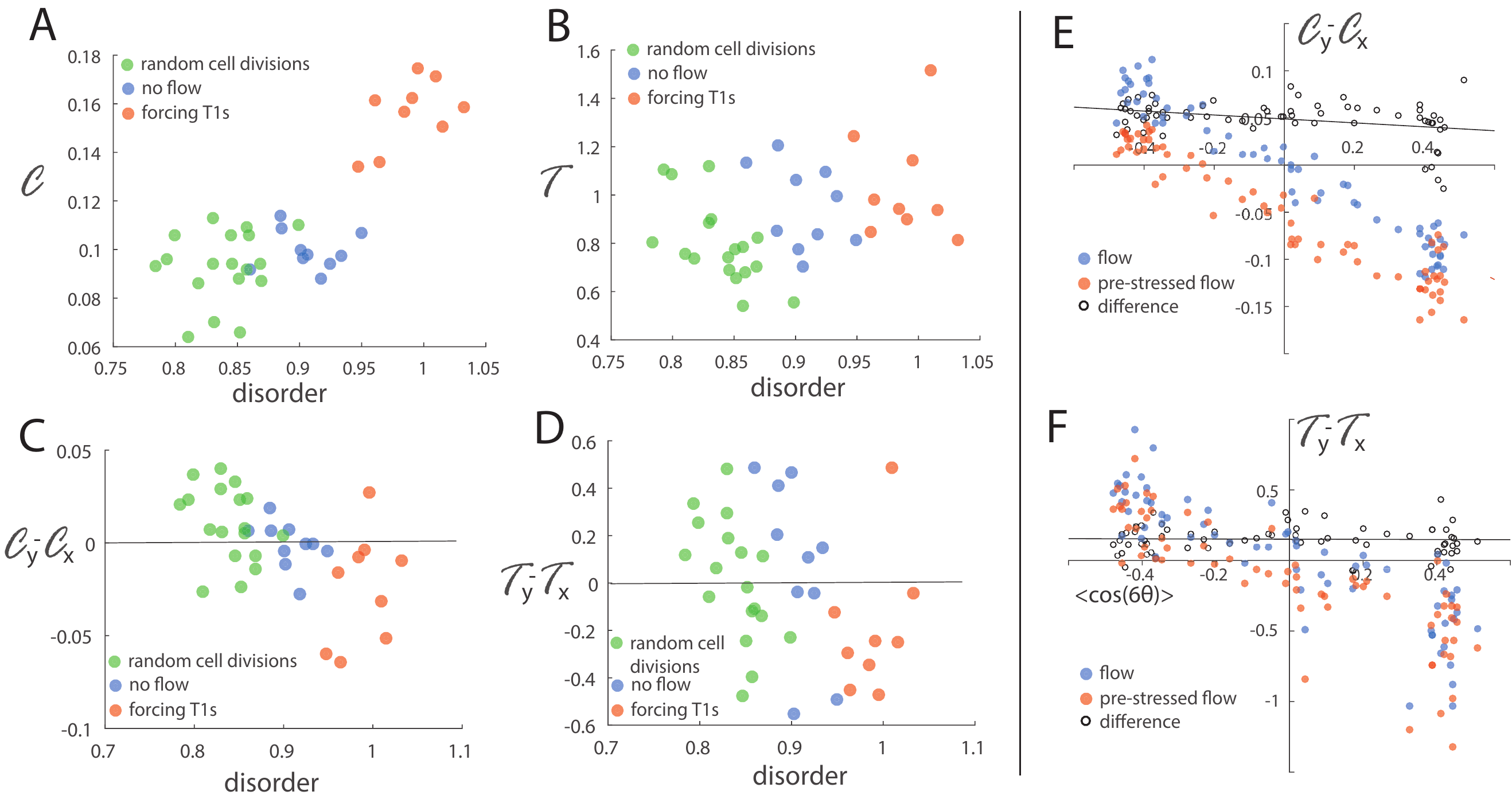}}
\caption{ {\bf Cableness and tissue disorder.} 
Input topologies with $\langle \cos(6\theta) \rangle \simeq 0$ were created through three different methods. In the first method cells underwent divisions at a random orientation. In the second method half of the vertical edges in an initially hexagonal packing were forced to undergo T1 transitions. In the final method the set of tissues with no flow (i.e. zero normalized T1 transitions) from Fig. 6 was used. The disorder on the horizontal axis ({\bf A}--{\bf D}) is defined as the standard deviation in edge number per cell throughout the tissue. Each point corresponds to one independently generated packing of 1000 cells. \\
{\bf A} $\mathcal{C}$ is strongly correlated with disorder. \\
{\bf B} $\mathcal{T}$ is more weakly correlated with disorder.  \\
{\bf C, D} Taking the difference in cableness between the $x$ and $y$ axis centers the data at 0 and removes most of the dependence on disorder. \\
{\bf E} The difference measure $\mathcal{C}_y-\mathcal{C}_x$ has the same trend between relaxed and pre-stressed tissues, with a vertical offset.    \\
{\bf F} The difference measure $\mathcal{T}_y-\mathcal{T}_x$ is decreased slightly by pre-stressing the tissue.    
}
\label{dis}
\end{adjustwidth*}
\end{figure}

We measured the cableness in these tissues and found that the values of $\mathcal{C}$ and $\mathcal{T}$ are linearly correlated with the edge number disorder (Fig. \ref{dis}). Increasing disorder decreases the cableness of tissues along a given axis. However, when we take the differences $\mathcal{C}_y-\mathcal{C}_x$ or $\mathcal{T}_y-\mathcal{T}_x$ in the cableness along the $y$ and $x$ axis we find that the difference does not depend on the disorder. Taking the difference in cableness along perpendicular axis has the additional benefit of centering the cableness of isotropic tissues at zero.We will use these modified measures as our definitions of cableness for the rest of the paper.  

%%%%%%%%%%%%%%%%%%%%%%%%%%%%%%%%%%%%%%%%%%%%%%%%%%%%%%%%%%%%%%%%%%%%%%%%%%%%%%%%%%%%%%%%%%%%%%%%%%%%%%%%%%%%%%%%%%%%%%%%%%%%%%%%%%%%%%%%%%%%%%%%%%%%%%%%%%%%%%%%%%%%%%%%%%%%%%%%%%%%%%%%%%%%%%%%%%%%%%%%%%%%%%%%%%%%%%%%%%%%%%%%%%%%%%%%%%%%%%%%%%%%%%%%%%%%%%%%%%%%%%%%%%%%%%%%%%%%%%%%%%%%%%%%%%%%%%%%%%%%%%%%%%%%%%%%%%%%%%%%%%%%%%%%%%%%%%%%%%%%%%%%%%%%%%%%%%%%%%%%%%%%%%%%%%%%%%%%%%%%%%%%%%%%%%%%%%%%%%%%%%%%%%%%%%%%%%%%%%%%%%%%%%%%%%%%%%%%%%%%
\subsubsection*{Validating cableness measures on pre-stressed tissues}
Up to this point we have always applied our cableness measures to stress free tissues. However, we would like to apply our measures to experimental data, which is frequently from tissues subject to applied stress. In order to verify that our measures hold on pre-stressed data we applied a stress anisotropy to the input topology packings generated by flow. We let the tensions on the edges evolve according to 
\begin{equation}
\beta_\gamma \dot{\gamma_i}= \big(\vec{F}_{a} - \vec{F}_b\big) \cdot{\hat{l}_i}, 
\end{equation} 
where $\vec{F}_a$ and $\vec{F}_b$ are the forces on the vertices of edge $i$ and $\beta_\gamma$ is a characteristic relaxation time. We did not allow any topological changes to occur. This process more closely resembles the behavior of cells under anisotropic stress than the stretching procedure which we use to determine cableness. Fig.~\ref{dis}E,F shows the results of applying our two cableness measures to both the pre-stressed and relaxed packings. Pre-stressing the tissue decreases the value of $\mathcal{C}_y-\mathcal{C}_x$ slightly without changing the overall trend, and had no noticeable effect on the tension based cableness measure. In introducing the idea of cableness we said that it is a property of a cell packing topology independent of its state of stress. These results show that our cableness measures satisfy this criterion to a good approximation.  (It is worth underlining that this is not true of all conceivable measures of cableness; for example, an intuitively appealing measure based on cell elongation which fails this test is described in S2 Appendix.)

%%%%%%%%%%%%%%%%%%%%%%%%%%%%%%%%%%%%%%%%%%%%%%%%%%%%%%%%%%%%%%%%%%%%%%%%%%%%%%%%%%%%%%%%%%%%%%%%%%%%%%%%%%%%%%%%%%%%%%%%%%%%%%%%%%%%%%%%%%%%%%%%%%%%%%%%%%%%%%%%%%%%%%%%%%%%%%%%%%%%%%%%%%%%%%%%%%%%%%%%%%%%%%%%%%%%%%%%%%%%%%%%%%%%%%%%%%%%%%%%%%%%%%%%%%%%%%%%%%%%%%%%%%%%%%%%%%%%%%%%%%%%%%%%%%%%%%%%%%%%%%%%%%%%%%%%%%%%%%%%%%%%%%%%%%%%%%%%%%%%%%%%%%%%%%%%%%%%%%%%%%%%%%%%%%%%%%%%%%%%%%%%%%%%%%%%%%%%%%%%%%%%%%%%%%%%%%%%%%%%%%%%%%%%%%%%%%%%%%%%%%%%%%%%%%%%%%%%%%%%%%%%%%%%%%%%%%%%%%%%%%%%%%%%%%%%%%%%%%%%%%%%%%%%%%%%%%%%
\subsection*{Oriented cell divisions promote cableness}
\begin{figure}[t]
\begin{adjustwidth*}{-2in}{-0in}
{\includegraphics[width=1.3\textwidth]{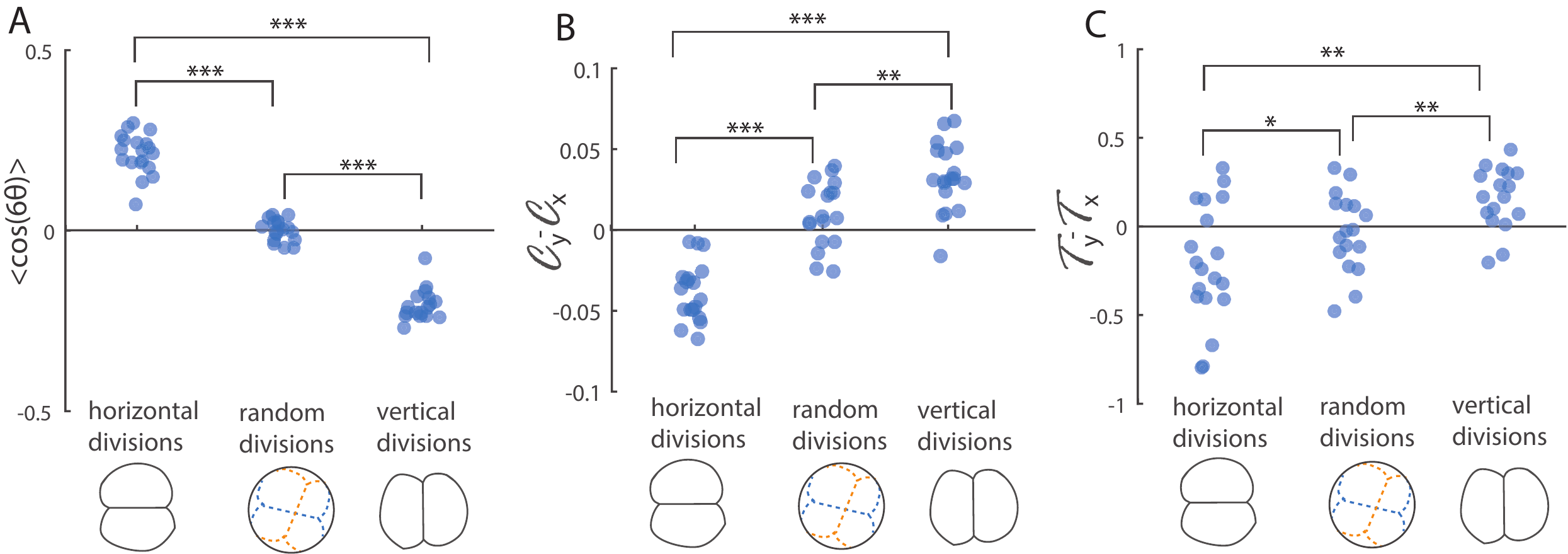}}
\caption{ {\bf Cableness of tissues with oriented divisions.} 
Each point corresponds to one packing topology with 1000 cells generated through cell divisions. Stars represent significance at the 0.05, 0.01, and 0.001 levels according to a t-test. \\
{\bf A} Edge orientation of packings generated by oriented cell divisions.  (Horizontal divisions are defined to be those with a horizontal cleavage plane, which would tend to be produced by a mechanical stress in the vertical direction.) \\
{\bf B} Result of applying the cell convexity based cableness measure $\mathcal{C}_y-\mathcal{C}_x$ to the packings generated by oriented cell divisions. \\
{\bf C} Result of applying the tension based cableness measure $\mathcal{T}_y-\mathcal{T}_x$ to the packings generated by oriented cell divisions. 
}
\label{OrientedDivisions}
\end{adjustwidth*}
\end{figure}

Given that passive cell flow in the direction of applied stress decreases the cableness of a tissue in that direction, we would like to find a fundamental topology-changing process that increases cableness. Here we show that oriented divisions are one such process. Elongated cells are known to divide preferentially perpendicular to their long axis which in turn tends to align with the applied stress \cite{LecuitDev2013, Bosveld2016, EatonSemcdb2017,KellyNatCom2017,HelenDev2014,Wyatt2015, Hertwig}. A cartoon of the vertex model used to create a statistical ensemble of packings derived from oriented divisions is given in Fig.~\ref{VMUsed} panel labeled `oriented divisions'. The simulation has periodic boundary conditions, constant edge tensions, and pressure forces from cells. The initial packing is a Voronoi tessellation of random points in the plane which has been isotropically relaxed allowing for topological changes. Every cell divides exactly once in a random order. For simplicity and consistency we directly impose the orientation of the cleavage plane. The cleavage plane is either horizontal, vertical, or at a random unbiased angle. It should be noted that this increases the internal stress in the tissue in agreement with \cite{ManiArxiv2018}.  The cableness of each tissue in this ensemble of packings was determined and the results are shown in Fig. \ref{OrientedDivisions}. Tissues in which the cells divide with a horizontal cleavage plane (as would be expected for a vertical applied stress) are more cably than tissues in which the cells divide vertically or in a random orientation.

\subsection*{Example: {\it Drosophila} epithelium}

\begin{figure}[ht!]
\begin{adjustwidth*}{-2in}{-0in}
{\includegraphics[width=1.2\textwidth]{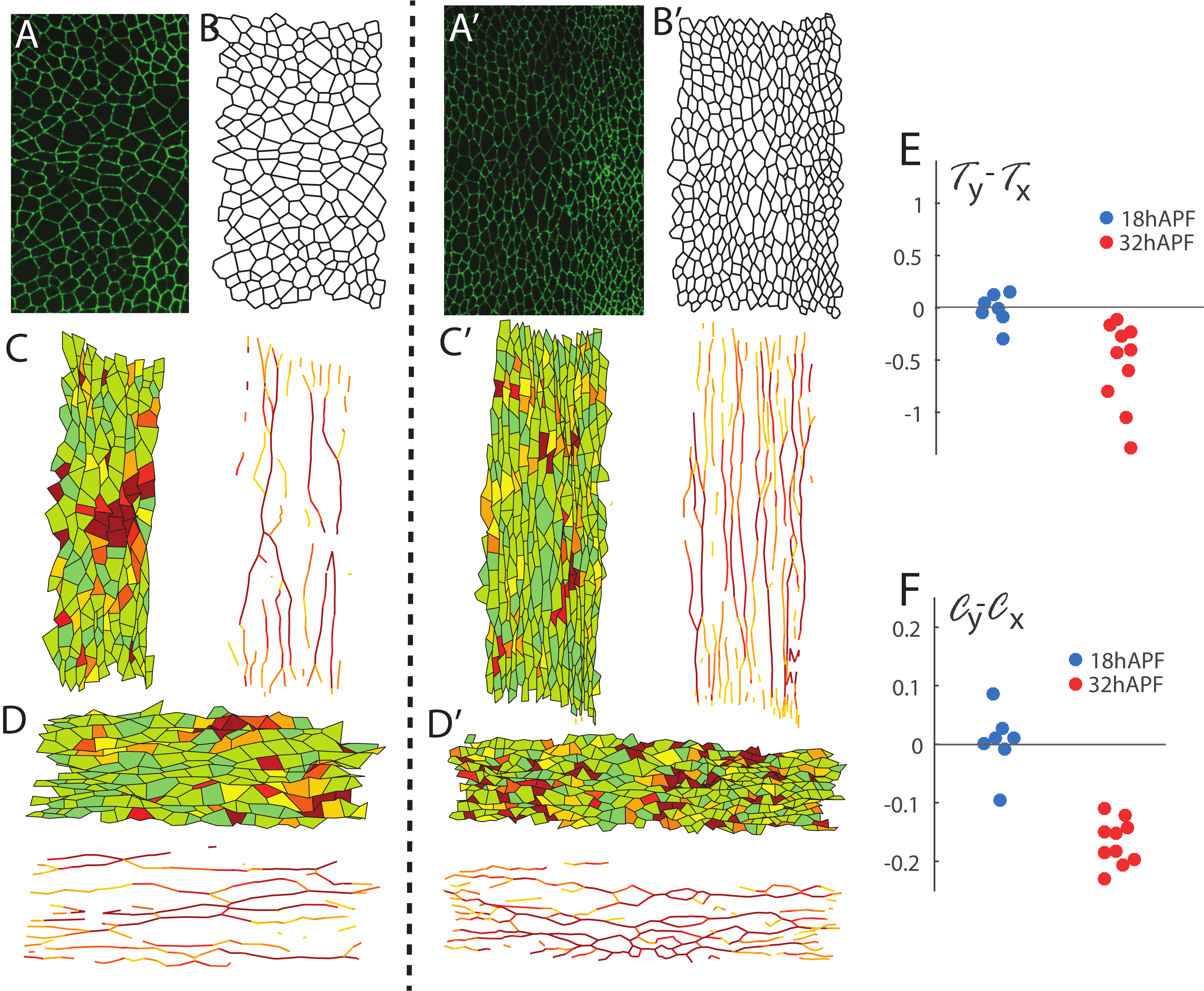}}
\caption{ {\bf Measuring cableness in the {\it Drosophila} pupa notum.}
Unprimed frames are taken at 18 hours after pupa formation. Primed frames are from a pupa imaged at 32 hours after pupa formation after the tissue has undergone a round of oriented cell divisions with (on average) a horizontal cleavage plane and is under significantly higher stress in the vertical direction. \\
{\bf A (A'):} Image of the apical cell outlines of the wild type tissue labeled by ECad-GFP. The midline runs horizontally through the center of the image, and posterior is to the left.   \\
{\bf B (B'):} Vertex model seeded from the pupa data. \\
{\bf C (C'):} Results of the pulling procedure along the vertical axis. \\
{\bf D (D'):} Results of the pulling procedure along the horizontal axis. \\
{\bf E:} Tension-based cableness measure. Each point represents one pupa. Tissues at 32 hours APF are more cably than tissues at 18 hours APF. \\
{\bf F:} Concavity based cableness measure. Each point represents one pupa. Tissues at 32 hours APF are more cably than tissues at 18 hours APF.
}
\label{pupa}
\end{adjustwidth*}
\end{figure}

In the previous two sections we have established several measures of cableness and shown that cableness increases  in the direction of applied tension when cells divide perpendicular to the applied tension. In this section we will give an example of how we can apply our measure to biological data. We looked at the {\it Drosophila} pupa notum at 18 and 32 hours after pupa formation (APF). This system is of interest because it is known to undergo a wave of oriented divisions along with an increase in stress anisotropy over this time period, leading us to hypothesize that the tissue should increase its cableness between 18 and 32 hAPF \cite{Guirao2015}. Fig.~\ref{pupa}A,B gives an example tissue at 18 and 32 hAPF. The images are skeletonized and used as the input topology for the stretching procedure. Image acquisition and segmentation were carried out as described in \cite{Guirao2015}. 

Fig.~\ref{pupa}C,D show the qualitative results of the stretching procedure applied both perpendicular and parallel to the midline. At 18 hAPF the tissue's response to the stretching procedure is the same along both the vertical and horizontal axis. By 32 hAPF the tissue's response to the stretching procedure is highly dependent on the axis of stretch. When stretched perpendicular to the midline most cells remain convex and edges form many cables which rarely branch. In contrast when stretched parallel to the AP axis the cells are forced into highly irregular shapes, and the lines of force through the tissue frequently branch. The quantitative cableness measures agree with our prediction that the tissue becomes more cably in the vertical direction in time. Thus in this system mechanical stress orients cell divisions and these divisions allow the system to become more cably in the direction of stress. This more cably tissue can then form actomyosin cables to oppose the applied stress and prevent further tissue elongation.

%%%%%%%%%%%%%%%%%%%%%%%%%%%%%%%%%%%%%%%%%%%%%%%%%%%%%%%%%%%%%%%%%%%%%%%%%%%%%%%%%%%%%%%%%%%%%%%%%%%%%%%%%%%%%%%%%%%%%%%%
%%%%%%%%%%%%%%%%%%%%%%%%%%%%%%%%%%%%%%%%%%%%%%%%%%%%%%%%%%%%%%%%%%%%%%%%%%%%%%%%%%%%%%%%%%%%%%%%%%%%%%%%%%%%%%%%%%%%%%%%%%%%
%%%%%%%%%%%%%%%%%%%%%%%%%%%%%%%%%%%%%%%%%%%%%%%%%%%%%%%%%%%%%%%%%%%%%%%%%%%%%%%%%%%%%%%%%%%%%%%%%%%%%%%%%%%%%%%%%%%%%%%%%%%%%%%%%%%%
%%%%%%%%%%%%%%%%%%%%%%%%%%%%%%%%%%%%%%%%%%%%%%%%%%%%%%%%%%%%%%%%%%%%%%%%%%%%%%%%%%%%%%%%%%%%%%%%%%%%%%%%%%%%%%%%%%%%%%%%%%%%%%%%%%%%
\section*{Conclusion}

Multiple parallel actomyosin cables can only form in tissues if they have a favorable topology, which we call cableness. In a cartoon model of a tissue, cableness corresponds to the average edge orientation as measured by $\langle \cos(6\theta) \rangle$. However, it is unclear whether $\langle \cos(6\theta) \rangle$ remains a good predictor of cableness in realistic, disordered tissues, and it is obviously of less use in tissues with few hexagonal cells, or in tissues in which the cells are elongated. Here we defined two cableness measures $\mathcal{C}_y-\mathcal{C}_x$ and $\mathcal{T}_y-\mathcal{T}_x$, which quantify intuitive notions of whether a tissue does a good job of forming cables. We find that these measures correlate well with $\langle \cos(6\theta) \rangle$ in unstressed, statistically isotropic tissues with low topological disorder, but also reflect a physically and biologically meaningful idea of cableness in more disordered or pre-stressed tissues.

The convexity measure $\mathcal{C}_y-\mathcal{C}_x$ describes the geometry of cably tissues. When a tissue has high cableness its cells form a brick-like structure, whereas tissues with low cableness tend to have highly elongated cells which become concave and overlap when disorder is introduced to the system. These properties may also shed light on the role of cableness in morphogenesis. We expect that tissues with high cableness will react better to anisotropic stresses, distributing the force evenly throughout the tissue while maintaining relatively round cells. 

The tension based measure $\mathcal{T}_y-\mathcal{T}_x$ describes the extent to which forces are spread evenly through the tissue. In disordered tissues that lack cableness only one or a few cables form in the tissue, whereas in tissues with high cableness many parallel cables form when the tissue is stressed. We expect that one could also define cableness measures based on the network properties of the stretched packings along the lines of network measures of force chains, and we hope that this work inspires additional work along these lines \cite{Miroslav2016, Giusti2016, Bassett2015}.

In order to maintain high cableness along one axis tissues must give up the ability to form cables along the perpendicular axis. However, tissues with a greater level of disorder also have lower cableness along a single axis. This suggests that cableness along a single axis is a function both of the orientational order of the cells and the total level of disorder in the tissue. 

Passive cell flow along the axis of higher stress in an epithelium has the effect of reducing the cableness in that direction. Unlike the well know alignment of liquid crystals due to flow, in this case the natural direction of the flow serves to disrupt the alignment of cells into a cably orientation \cite{LarsonChemPhy1995,Schmidt1995,Hess1999}. Sugimura and Ishihara made note of this in the \textit{Drosophila} pupal wing from around 24 to 32 hours after pupa formation. During this time there is cell flow along the proximal-distal (PD) axis and the cells become less cably in that direction as measured by $\langle\cos(6\theta)\rangle$ of hexagonal cells \cite{IshiharaDev2013}. Therefore, the only way for cells to align in the CFO through T1 transitions is if the T1 transitions shrink edges parallel to the high stress axis. This requires some amount of overshoot in the tension on these edges, so that they constrict under stress rather than elongating. This appears to happen in the \textit{Drosophila} pupal wing prior to cell flow \cite{EatonSemcdb2017} and in the \textit{Drosophila} notum \cite{Guirao2015}. When cell flow is driven by internal stress we predict that cables will collapse into multi-cellular rosettes as seen in \textit{Drosophila} intercalating cells \cite{ZallenDevCell2009}.

A second process that increases the cableness of a tissue is oriented cell divisions. It has been well established for centuries that cells tend to divide along their long axis \cite{Hertwig, HelenDev2014}. Therefore, in a tissue under high vertical stress, cells will elongate vertically and tend to divide with a horizontal cleavage plane, thus increasing the cableness of the tissue in the vertical direction. It has previously been argued that oriented divisions can relax stress by elongating tissue in the direction of the imposed pulling \cite{Wyatt2015,Campinho2013}.  Here, we report a complementary role: they also cause packing topology to change so that cells are better able to form oriented cables and resist deformation by external stress. Both multicellular myosin cables and oriented divisions along the high stress axis are seen in tissues including the \textit{Drosophila} larval wing and the mouse heart \cite{TaponEmbo2013, LecuitDev2013, KellyNatCom2017}. 

Studies of multicellular actomyosin cables have to date focused largely on isolated cables that form at well-defined boundaries, for example between two different tissues or compartments. Here we hope to spark interest in the role of in morphogenesis of collections of parallel cables within a single tissue.  Whereas a single cable can form in almost any arrangement of cells, we have argued here that arrays of parallel cables are feasible only in cell packings that are sufficiently cably. Moreover, because the natural cell flow in the direction of applied stress tends to 'undo' cableness, anywhere that multiple cables are seen one must ask the question: how did the cells arrange themselves to allow cables to form? Here we have shown that one natural mechanism for increasing a tissue's ability to form cables is oriented cell divisions. 

We have proposed two initial ways of quantifying a tissue's ability to form cables. While these methods have many promising features, we believe that further study of cables, perhaps from a network standpoint, may yield even better ways of quantifying cableness. We have argued that any measure of cableness must work on both relaxed and stressed tissues since cableness, as we define it, is a property of a cell packing topology, not a measure of edge alignment. 

\section*{Acknowledgments}
This material is based upon work supported by the National Science Foundation under Grant No. DMR-1056456, DGE-1256260 and IOS-1353914 as well as the Margaret and Herman Sokol Faculty Awards.

\end{document}